# Increasing Network Visibility Using Coded Repetition Beacon Piggybacking

Ghassan Samara[1], Wafaa A.H. Ali Alsalihy[2], Sureswaran Ramadass[3]

[1]National Advanced IPv6 Centre, Universiti Sains Malaysia, 11800 Penang, Malaysia
[2]School of Computer Science, Universiti Sains Malaysia, 11800 Penang, Malaysia

*Abstract*—Vehicular Ad hoc Networks (VANET) is one of the most challenging research areas in the field of Mobile Ad Hoc Networks. In this research, we propose a new mechanism for increasing network visibility, by taking the information gained from periodic safety messages (beacons), and inserting it into a 'neighbor' table. The table will be propagated to all neighbors giving a wider vision for each vehicle belonging to the network. It will also decrease the risk of collision at road junctions as each vehicle will have prior knowledge oncoming vehicles before reaching the junction.

## I. INTRODUCTION

VANET has attracted a wide range of research effort these days, aiming to reach road safety, infotainment and a comfortable driving experience. Beacon message sometimes called status messages as they give information about the status of the sender need to be propagated frequently over a wide range, and for a large number neighboring vehicles, to give information about current vehicle status. To make vehicles aware of the status of the whole network, this critical information must be sent with high probability and reliability.

Normally in VANET each vehicle transmits its information periodically to its neighboring vehicles. This information transmitted via periodic safety messages called beacons; beacon contains vehicle status and real time information, see Fig. 2. VANET uses special radios for transmission called Dedicated Short Range Communication (DSRC) radios, which guarantees to reach a distance up to 300m when sending a beacon in best conditions. Beacon helps vehicles and drivers by providing critical information and gives a prior knowledge that will help to avoid danger before it is reached.

The IEEE 802.11p [1] spectrum band has allocated seven 10MHz channel one of which is a control channel which is used for safety and control messages. Utilizing this channel must be done wisely as it is the only channel for this kind of messages and all vehicles must compete to use it.

In this paper, we are concerned with distributing beacon information, especially the information about position, speed, and direction. This provides vehicles in the network extended information about the image of current networks.

We will explore many cases for sending information in normal freeways and junctions. To the best of our knowledge, sending beacon information in junctions has not been explored before. In sec. 2, we analyze the current research efforts in the area of safety-message transmission of VANET. In sec. 3, we address our proposed network that contains solutions for the current system. In sec. 4, we show the result of our simulation.

## II. ANALYSIS OF RELEVANT RESEARCH AREA

Many papers have introduced the idea of how to increase the reception of beacons. In [2], the authors proposed an idea of piggybacking the neighbor beacon to increase the safety message reception in freeway But, piggybacking beacon itself will double the original beacon size. Authors in [3] presented the idea of ensuring the successful reception of beacons by exchanging the power information of each vehicle using beacon piggybacking.

Another idea, in [4], proposed a coded-cooperative-repetition scheme for safety message broadcasting. As two beacons are coded into one packet, this technique which is called CCB protocol requires sending additional packets (beyond the original beacons), and the result will not give a wide vision about the network.

Trying to send beacons aggressively to the entire network will increase 'collisions'. The idea of resending a received beacon to the entire network causes the 'broadcast-storm' problem [5].

In [6], authors used a neighbor table to know the neighbors of each vehicle to decide the next forwarder, this technique is used for the greedy forwarding scheme, in [7] each node examining about the correctness of neighbor position to avoid receiving falsified data, by enabling each node to use multiple sensors to detect malicious or selfish behavior of nodes in the network and store these data in neighbor table.

Authors in [8] presented analytical methods to study beaconing in VANET. Numerical results demonstrate that these methods could be used to estimate the probability of successful beacon delivery and mean beacon transmission delay to check the car-to-car application requirements.

Xu et al. [9] first exploited the repetition mechanisms. Multiple retransmissions of the same message are performed by the same source vehicle to overcome channel failures and message collisions. Although the evaluation results confirmed this method's potential, room was left for improvements.

### III. PROPOSED NETWORK

*A. Basic Idea*

Each vehicle transmits a status message called beacon every 100 ms [10] and at the same time it receives beacons from its neighbors, this beacon contains ID, position, direction, speed, time stamp, and beacon interval [11]. Each vehicle is equipped with a GPS device to obtain the current position [12].

*1) Preparing to send*

After receiving beacons from its neighbors, a vehicle starts to collect the information gained and inserts it into Neighbor Table (NT). NT contains the ID, position, speed and direction of each sending vehicle. The information inside the table is ordered so that neighbors close to the vehicle will be at the top. This helps the vehicle to draw the topology of the network around it. NT is updated every 1 second if the channel is not congested. The method of computing channel congestion, presented in (19) and in (20), is based on the beacons received in one second by the vehicle, see fig 1. From the figure we can see that this vehicle received 8 beacons from Vehicle A for example and 6 beacons from Vehicle B and so on. To compute the channel congestion see (1).

| Vehicle A | 15 | 16 | 17 | 18 |    | 20 | 21 |    | 23 | 24 |
|-----------|----|----|----|----|----|----|----|----|----|----|
| Vehicle B | 71 | 72 |    |    | 75 |    |    | 78 | 79 | 80 |
| Vehicle C | 89 | 90 |    |    |    |    |    | 96 | 97 |    |
| Vehicle D | 22 | 23 | 24 | 25 | 26 | 27 |    |    | 29 | 30 |
| Vehicle E | 61 | 62 | 63 |    |    |    | 67 |    | 69 | 70 |

**Fig. 1 Beacons received by a vehicle in 1 second(19), (20).**

$$CP = (1 - \frac{\sum B}{N \times 10}) \times 100\%, \qquad (1)$$

where CP is the congestion probability, B is beacon received, and N is number of neighboring vehicles.

Every second, the vehicle evaluates the channel congestion to examine the channel status. If CP < 50%, this means that the channel is not highly congested and the vehicle can proceed to build NT to be transmitted to neighbors. If CP>50, then a vehicle will not build NT to avoid further congestion; as building NT and piggybacking it to a current beacon would only increase the channel congestion. In this case vehicle will wait for a second before checking the channel congestion again.

Building the NT starts when a vehicle inserts the ID, position, speed, and direction of every neighbor. After making the NT, a vehicle assigns a Time Stamp (TS) and attaches the Life Time (LT) of NT that expires after 1 second, as after 1 second the information will be too old to be meaningful. It also attaches the Sequence Number (SN) taken from the MAC layer, according to IEEE 802.11 [13] standards. A two-byte sequence control field is contained in an 802.11 MAC header- sequence number that assures this NT has not been used or received before.

| ts | bi | pos | sp | $H_1$.pos | $H_1$.sp | $H_2$.pos | $H_2$.sp | … | u |
|----|----|----|----|----|----|----|----|----|----|

ts: timestamp  
pos: position of x  
$H_i$.pos: header of lane i position  
$H_i$.sp : header of lane i speed  
bi: beacon interval  
sp: speed of x  
u: unused

**Fig. 2: Beacon illustration [11].**

Sending Steps:

1. Vehicle receives beacons from its neighbors periodically, up to 10 beacons from each neighbor every second.
2. Vehicle computes the channel congestion.
3. Vehicle decides whether to build NT or not, depending on CP. If CP<50 %.
   a. Vehicle updates its NT, by taking the latest information gained from the beacons arrived.
   b. Insert ID, Position, speed and direction into NT.
   c. Rearrange the table; put the nearest vehicles at the top of the table.
   d. Vehicle assigns a SN and life time for NT.
   e. Vehicle piggybacks NT on beacon.
   f. Vehicle sends the beacon.
4. If CP≥50%
   a. Wait 1 seconds
   b. Go to 2

See algorithm 1 for sending beacon.

This algorithm lists the steps for preparing the beacon before sending it; it starts with receiving beacons from other vehicles, and clearing NT every 1 second to include the freshest information taken from beacons. In step 3 vehicle starts to add all beacons recently received from neighbors to NT if CP<50 %. After preparing the NT, a vehicle assigns some parameters to help the receiver to analyze the received beacon efficiently.

*2) Receiving a Beacon*

To gain information in a beacon from another vehicle requires many steps. The received beacon provides fresh information about the sender, so this information must be taken into consideration.

Coded Repetition Neighbor Table (CRNT): the received beacons from neighbor vehicles don't provide an image for the whole topology of the network, as many vehicles are not seen or are out of coverage. To increase the awareness of the network, we are proposing the use of a coded repetition technique. The technique starts when a vehicle senses the channel status and decides to build the NT and piggyback it every one second to a beacon. It sends this beacon to all the neighboring vehicles and, at the same time, it receives many beacons that contain the piggybacked NT (PNT) from neighboring vehicles. Piggybacking NT to the current beacon will not cause collision in the network as no additional beacons will be sent. The piggyback solution is proposed before in VANET in [2], [4], [3], [11], [14], [15], [16].

```
1.   // Sending Piggyback Beacon Algorithm
2.   Receive Beacon
3.   //Timer works every 1 second
4.   Timer()
5.   {
6.   Clear NT
7.   cp = (1 - ΣB/(N×10)) × 100 %//Compute channel congestion
8.   If Cp < 50%
9.   {
10.  For int i=0 to No. of Neighbors -1
11.  {
12.  Insert ID(i) into NT
13.  Insert Position(i)
14.  Insert Speed(i)
15.  Insert Direction(i)
16.  } // end for statement
17.  Rearrange descending according to position
18.  Assign TS
19.  LT = TS + 3
20.  Attach LT
21.  Attach SN
22.  Piggyback to a beacon
23.  } //end if
24.  Send beacon
25.  } // end of Timer
```

Algorithm 1: Sending Piggyback Beacon

```
// Receiving Beacon Algorithm
1.   Receive beacon
2.   If Beacon<-PNT = True
3.   {
4.     Extract PNT
5.     For int i=0 to SL length -1
6.     {
7.       If SL<-ID(i) = ID
8.       {
9.         If SL<- SN[i] >= SN
10.        {
11.        Move SL<-SN[i] to top
12.        End
13.        }// end of line 9 if
14.        Else
15.        {
16.        If LT < Current Time
17.        {
18.        CRNT = NT U PNT
19.        Add SN to top of SL
20.        Add Receive Time (RT) to SL
21.        }// end of line 16 if
22.        }// end of line 14 else
23.        }// end of line 7 if
24.      }// end of line 5 for
25.   }// end of line 2 if
```

Algorithm 2: Receiving Beacon

When any vehicle receives a beacon, it examines the beacon and checks whether the beacon contains a PNT or not. If it doesn't have one, then it will consider this beacon as a normal safety message. If it has one, the vehicle will extract the PNT and check to see if it been received before. To do this, each vehicle must refer to its Sequence List (SL). This list contains the sequence number of each neighbor vehicle that has sent a PNT before, the ID of that vehicle, and the time of receive. After extracting the PNT, a vehicle takes the ID of PNT sender and searches for it in the SL. If it is there, vehicle will compare the sequence number for the matched ID. If the sequence number of PNT is less than the SN of SL, this means that this PNT is old and there is no need to explore it again. If SN is greater, it will check for the lifetime of PNT. If not expired, vehicle will accept PNT for coding.

The coding technique will be performed by UNION operation for current CRNT with PNT and the result will be inserted into the CRNT. CRNT will give the vehicle extended information about the current topology of the network, as the vehicle will now have the latest information about all of its neighbors.

Steps for Receiving:
1. Receive beacons.
2. Check if the received beacon contains PNT or not.
3. If the beacon contains PNT, extract PNT.
4. Vehicle checks if PNT sequence number is old, and received before, and checks if the life time is expired or not.
5. If SN is new and life time is not expired:
   a. Vehicle makes Union between CRNT and PNT.
   b. Add ID, SN and time of the operation of PNT to Sequence List.

See algorithm 2 for receiving beacon. This algorithm describes the steps taken for receiving a beacon. As not all beacons contain a PNT, each beacon must be examined to know if it contains PNT or not. If the vehicle does this all the time, this will cause processing overhead. To overcome this problem the receiver will not test the arrived beacons from a vehicle for 99ms after receiving beacon has PNT. In other words, after receiving a beacon has PNT from a specific vehicle, the receiver will not test the rest of beacons coming from that vehicle for 99ms. After extracting the PNT, the receiver searches in its sequence list (line 5) to see if this vehicle has sent a PNT before (line 7). If this PNT was not received before (line 9) and if its life time had not expired, then do the union operation.

*3) CRNT Cases*

Our proposed technique can solve many cases in normal freeway traffic or at junctions, which are considered hard problems to transmit or receive a safety message.

*a)    Freeway:*

In straight roads, vehicles can gain information easily, but in some situations. Vehicle can't see all vehicles on the road; our mechanism aims to extend the vehicles knowledge for greater distances. See, for example, Fig. 2 and table 1.

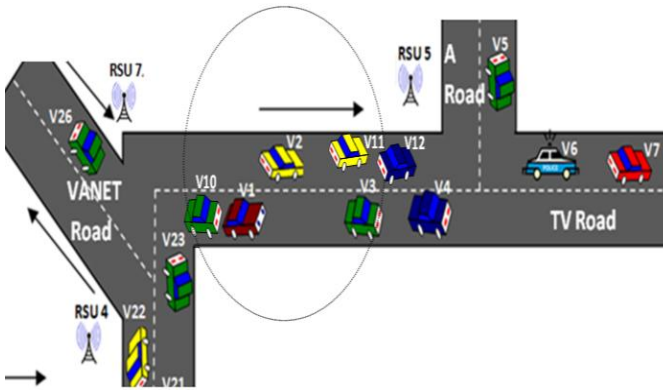

Fig. 2: Free Way

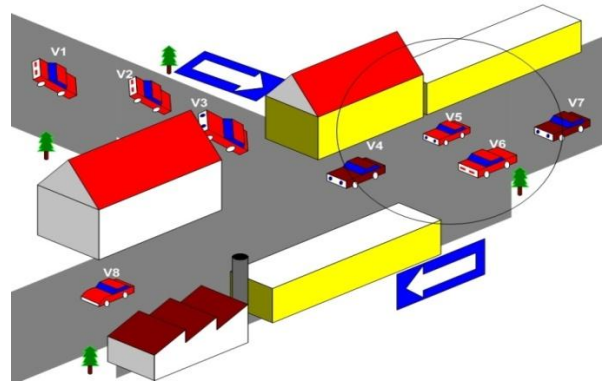

Fig. 3 a: Cross or T Junction, showing the communication range for V5

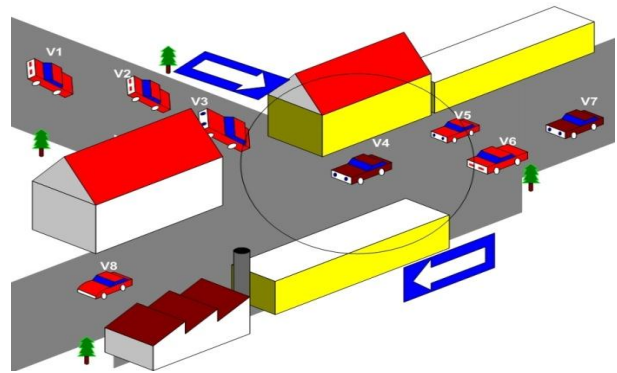

Fig. 3 b: Cross or T Junction, showing the communication range for V4

*b)   Cross or T Junction:*

Figure 3 is a case were vehicles are approaching a cross junction or T junction and don't have information about the incoming vehicles from the intersecting road, as buildings or other obstacles are blocking the beacons. In our example, V5 is approaching the junction and doesn't know that V3 is coming; V4 knows about V3 and broadcasts this information to its neighbors. Tables 3 and 4 represent CRNT for V5 and V4. Vehicles like V3 also must have information about the incoming road, and it must broadcast it to other neighbors, see table 5.

TABLE 1: CRNT FOR V2

| V2 | | | |
|---|---|---|---|
| **ID** | **Position(Lat, Long)** | **Speed** | **Direction** |
| V11 | 0.094449823, 1.751159661 | 30 | E |
| V3 | 0.094449823, 1.761159661 | 40 | W |
| V1 | 5.924449823, 7.51959661 | 20 | W |
| V10 | 0.094449823, 1.755159661 | 10 | W |
| V5 | 0.095449823, 1.755159661 | 20 | S |
| V4 | 5.924441823, 7.51159661 | 40 | W |
| V23 | 0.094049823, 1.755159661 | 30 | S |
| V22 | 0.094479823, 1.755159661 | 30 | N |
| V26 | 0.09444000, 1.755159661 | 20 | SE |

TABLE 2: CRNT FOR 10

| V10 | | | |
|---|---|---|---|
| **ID** | **Position(Lat, Long)** | **Speed** | **Direction** |
| V1 | 5.924449823, 7.51959661 | 20 | W |
| V2 | 5.924449823, 6.51759661 | 20 | E |
| V23 | 0.094049823, 1.755159661 | 30 | S |
| V22 | 0.094479823, 1.755159661 | 30 | N |
| V26 | 0.09444000, 1.755159661 | 20 | SE |
| V11 | 5.924419823, 6.51159661 | 30 | E |
| V3 | 7.924442823, 3.51159661 | 40 | W |
| V21 | 0.124449023, 11.51159661 | 50 | S |

Table 1 represents the CRNT for vehicle 2. As V3, V11, V1, V10 and V12 are neighbors, V2 can send and receive from them. The rest of the vehicles (in red) are neighbors of V3, V11, V1, V10 and V12, so V2 can know about V22 for instance, which is far from it (about 500m).

TABLE 3: CRNT FOR V5

| V5 | | | |
|---|---|---|---|
| **ID** | **Position(Lat, Long)** | **Speed** | **Direction** |
| V4 | 5.924441823, 7.51159661 | 20 | W |
| V6 | 5.924449823, 7.71159661 | 30 | W |
| V7 | 6.924499823, 4.51159661 | 40 | W |
| V3 | 7.924442823, 3.51159661 | 20 | S |

TABLE 4: CRNT FOR V4

| V4 | | | |
|---|---|---|---|
| ID | Position(Lat, Long) | Speed | Direction |
| V6 | 5.924449823, 7.71159661 | 30 | W |
| V5 | 0.095449823, 1.755159661 | 30 | W |
| V3 | 7.924442823, 3.51159661 | 20 | S |
| V7 | 6.924499823, 4.51159661 | 40 | W |
| V2 | 5.924449823, 6.51759661 | 30 | S |

TABLE 5: CRNT FOR V3

| V3 | | | |
|---|---|---|---|
| ID | Position(Lat, Long) | Speed | Direction |
| V4 | 5.924441823, 7.51159661 | 20 | W |
| V2 | 5.924449823, 6.51759661 | 30 | S |
| V5 | 0.095449823, 1.755159661 | 30 | W |
| V6 | 5.924449823, 7.71159661 | 30 | W |
| V1 | 5.924449823, 7.51959661 | 40 | S |

TABLE 6: CRNT FOR V9

| V9 | | | |
|---|---|---|---|
| ID | Position(Lat, Long) | Speed | Direction |
| V8 | 5.924449823, 6.51159661 | 30 | NW |
| V10 | 5.929449823, 6.51159661 | 30 | NW |
| V1 | 5.924449823, 7.51959661 | 20 | NW |
| V11 | 5.924419823, 6.51159661 | 40 | N |
| V2 | 5.924449823, 6.51759661 | 30 | N |

TABLE 7: CRNT FOR V8

| V8 | | | |
|---|---|---|---|
| ID | Position(Lat, Long) | Speed | Direction |
| V9 | 5.924649823, 6.58159661 | 30 | NW |
| V10 | 5.929449823, 6.51159661 | 30 | NW |
| V1 | 5.924449823, 7.51959661 | 20 | NW |
| V11 | 5.924419823, 6.51159661 | 40 | N |
| V2 | 5.924449823, 6.51759661 | 30 | N |

TABLE 8: CRNT FOR V1

| V1 | | | |
|---|---|---|---|
| ID | Position | Speed | Direction |
| V8 | 5.924449823, 6.51159661 | 30 | NW |
| V9 | 5.924649823, 6.58159661 | 30 | NW |
| V11 | 5.924419823, 6.51159661 | 40 | N |
| V2 | 5.924449823, 6.51759661 | 30 | N |
| V10 | 5.929449823, 6.51159661 | 30 | NW |
| V3 | 7.924442823, 3.51159661 | 50 | S |
| V4 | 5.924441823, 7.51159661 | 30 | N |
| V6 | 5.924449823, 7.71159661 | 50 | S |

*c. High Speed Junction:*

Where a vehicle wants to enter a highway without prior knowledge about it, see figure 4.

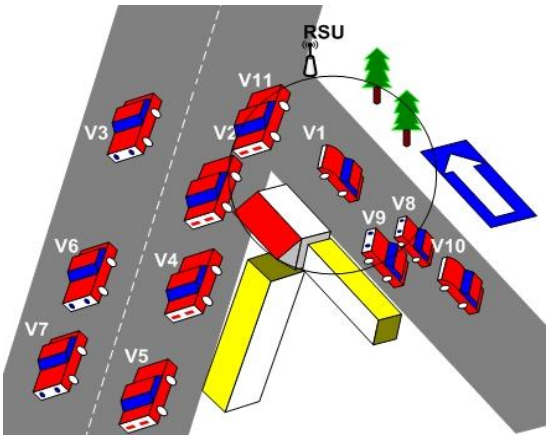

Fig. 4: High Speed Junction

The difference between this case and the previous case is that the vehicles in the highway have high speed.

In this example, V1 knows about V11 and V2; but V9 and V8 do not have that knowledge directly. However, they can take it from V1, see CRNT for V9, V8 and V1 in tables 6, 7 and 8.

*C. Other Considerations*

*1) Sparse Area*

Sparse area is an area were a vehicle doesn't receive any beacon from neighbor vehicles, so NT is empty and there is no need to transmit it to other vehicles. In this case, the vehicle will transmit its information to any Road Side Unit (RSU) at road junctions [17] to inform incoming vehicles about this vehicle.

*2) Tradeoffs:*

- Size of NT is small, as it only contains information just about vehicle position, speed, and direction. If it contains more information, it will be bigger, and would cause network overhead by sending a large beacon every second.

- Data collected only from neighbors represent first level data. If extended for another level of neighbors, it will increase overhead, and will be old, as the aggregation need to be processed in each vehicle before being retransmitted again.

- Frequency of transmission is once every second. If it is higher, it will cause network overhead. If it is lower, the data will be old, or too late for proper response. Since this information is about safety, it must be up to date and transmitted in a reasonable time.

## IV. SIMULATION AND RESULTS

### A. Simulation Setup

In order to test correctness of our protocol we made the simulation using the commercial program Matlab®, we have selected the use of Matlab as the version R2010b provides a complete and almost real environment for VANET, Matlab dedicated CAN tool (14) that can simulate the VANET channel by using Kvaser and Vector Drivers (22), we have created the messages and signals using CANoe Tool (23) that is dedicated to manage the DBC (Database) files, we also used AWGN channel to add noise to the signal, the distribution used is Nakagami distribution model [18] with fading intensity = 3 as suggested in [18].

| Parameter | Value |
|---|---|
| Radio propagation model | Nakagami-m, m = 3 |
| IEEE 802.11p data rate | 6Mbps |
| Bus Speed | 10000 |
| PLCP header length | 8 μs |
| Symbol duration | 8 μs |
| Noise floor | -99dBm |
| SINR | 10 dB |
| CW Min | 15 μs |
| CW Max | 1023 μs |
| Slot time | 16 μs |
| SIFS time | 32 μs |
| DIFS time | 64 μs |
| Message size | 512 bytes |
| CAN Device | Kvaser Virtual |
| Periodic Message Rate | 10 Message / s |
| Number of Cars | 200 |
| Road Length | 2 KM |
| Car Speed | 20km – 120km |
| Simulation Time | 10 s |
| Potential forwarders threshold | 10 |
| Communication Range | 300m |

TABLE 8: SIMULATION CONFIGURATION VALUES

Parameters used in our simulation are summarized in table 3; all the simulations in this paper will adopt these parameters.

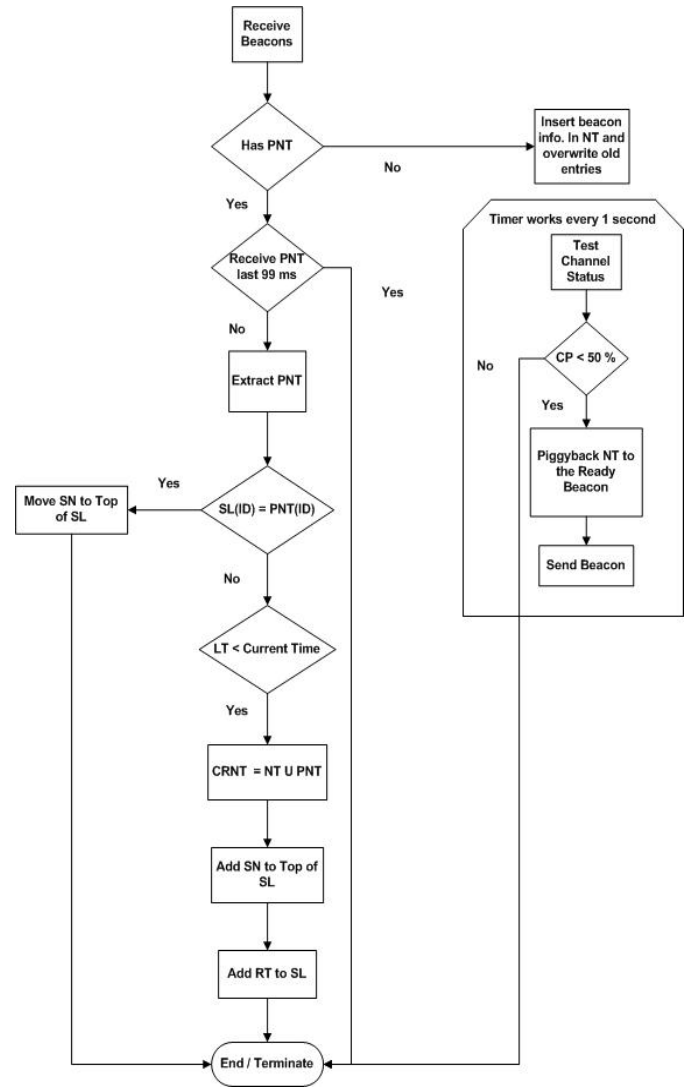

Fig. 5: Data Flow Diagram for the proposed system.

We made our simulation for 10s including 200 vehicles in 2km road consisting of 3 lanes.
The result could be seen in figures 6, 7, 8 and 9.

### B. Simulation Results

In our simulation we have tested the performance of our protocol CRNT and compared it with the performance of the CCB protocol (4) on highway scenario as it is the only scenario available for CCB to compare with.

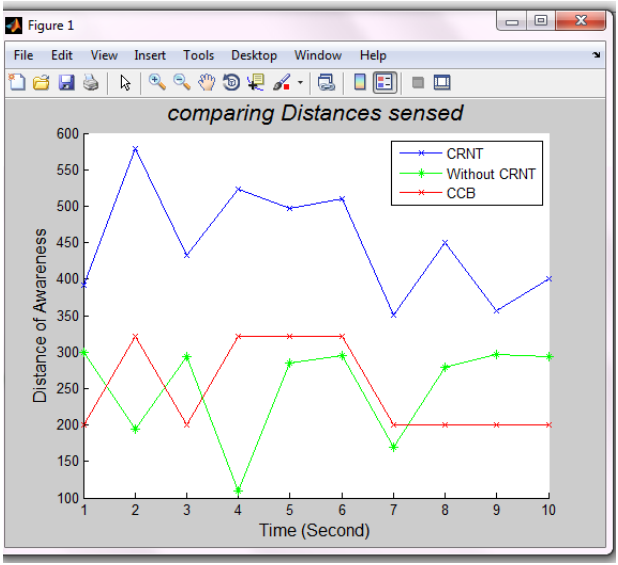

Fig. 6: Comparing Distance Sensed by a vehicle for CRNT and CCB protocols.

In figure 6 which compares distance of awareness provided by the CRNT and CCB protocols, the lines represent the distance of vehicle that could be seen. Beacons can reach 300m in best cases; but for most channel and network conditions, this number becomes a challenge to achieve. In this figure, the maximum distance for reception was 300m for without using the protocols, 340m achieved by CCB. These numbers are greatly improved by our technique, which as seen in the figures, increases the visibility by more than 570m in some cases.

In some cases the visibility is very limited, for instance in the second 4, where NT gives information about neighboring vehicles for away 100m, CCB gives information for 320, while CRNT for 530, this means that even though the reception of beacons is decreased, CRNT still gives increasing information about the vehicles in the network.

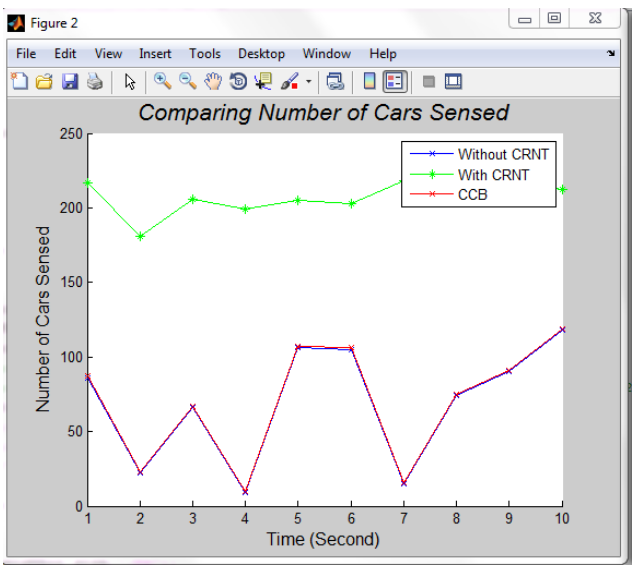

Fig. 7: Comparing Number of Cars Sensed by a vehicle for CRNT and CCB protocols.

The number of vehicles provided by CRNT is greater than the number of vehicles provided directly from beacons or from CCB. In this figure, we can see that the number of vehicles is greater using CRNT than the number gained from beacons and CCB and these numbers of vehicles monitored and range of visibility is approximately doubled.

We can see also that, even if the vehicle has low beacon reception from neighbors, it can have wide range information taken from CRNT about neighbors.

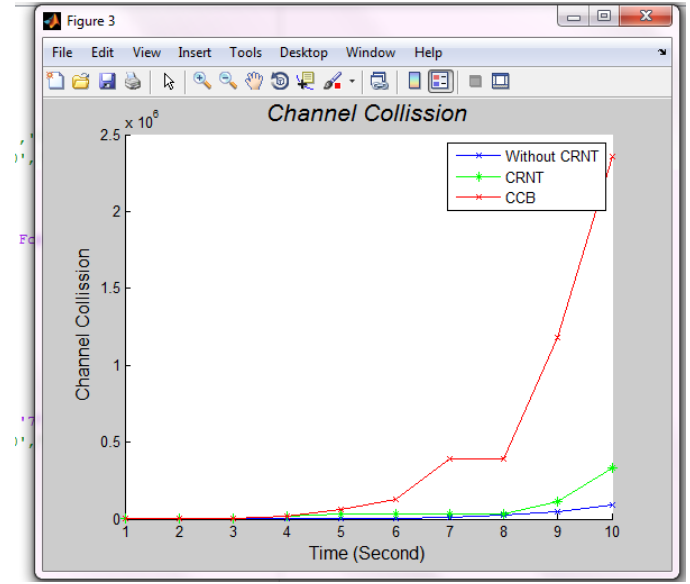

Fig. 8: Comparing the collision produced on the channel by using CRNT and CCB protocols.

The collision produced by CCB is increasing over time, CRNT scores less collision than CCB as it is been transmitted only if the channel is not congested, we can see that CRNT causes higher collision than beacon but this collision percentage still reasonable.

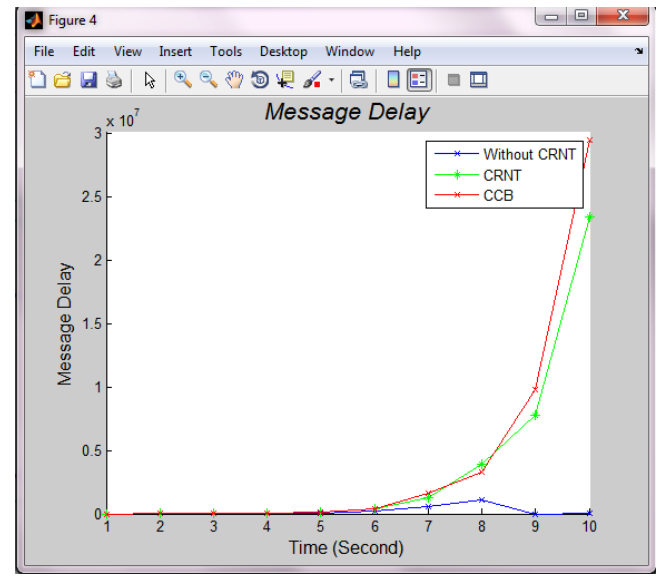

Fig. 9: Comparing the message delay when sending message using CRNT and CCB protocols.

The message delay for CRNT is shorter than CCB, when CRNT scores shorter message delay means that a rich information received by neighboring vehicles with short delay.

## CONCLUSION

In this research we conducted an extensive simulation study in order to evaluate the performance of information gathered from beacons in vehicular ad hoc networks. Taking into consideration that the beacon can't reach more than 300m and this will give a limited visibility for vehicles in the network, this issue is a hot topic for research. We realized that the visibility gained from a beacon alone is not enough; so, in order to improve the performance, we introduced a new technique for increasing the visibility of the network using the Coded Repetition Beacon Piggybacking on neighbor tables received from neighbors. Our simulation shows that this technique can help to overcome the shortage of vision for each vehicle, also CRNT produces low and reasonable collision compared with other protocols and short message delay.

## ACKNOWLEDGMENT

This paper is financially supported by Institute of Postgraduate Studies (IPS) Fellowship Scheme, Universiti Sains Malaysia (USM).